    \documentclass[aps,draft,twocolumn,eqsecnum]{revtex4}

\begin{document}

\title{Constraint propagation in $(N+1)$-dimensional space-time}
\author{Hisa-aki Shinkai} \email{shinkai@einstein1905.info}
\author{Gen Yoneda} \email{yoneda@waseda.jp}
\affiliation{
${}^\ast$ Computational Astrophysics Lab.,
Institute of Physical \& Chemical Research (RIKEN), \\
Hirosawa, Wako, Saitama, 351-0198 Japan \\
(Present Address: Hanayama 1-41, Nagaoka-kyo, Kyoto 617-0842 Japan)
\\ ~\\
${}^\dagger$ Department of Mathematical Science, Waseda University,
Okubo, Shinjuku, Tokyo,  169-8555, Japan
}

 \date{September 25, 2003 (submitted), to appear in Gen. Rel. Grav. (2004)}
\begin{abstract}

Higher dimensional space-time models provide us an alternative
interpretation of nature,  and give us different dynamical aspects
than the traditional four-dimensional space-time models.
Motivated by such recent interests, especially for
future numerical research of higher-dimensional space-time,
we study the dimensional dependence of constraint propagation
behavior.
The $N+1$ Arnowitt-Deser-Misner evolution
equation has matter terms which depend on $N$,
but the constraints and constraint propagation equations remain the
same.  This indicates that there would be problems with accuracy and
stability when we directly apply the $N+1$ ADM formulation to numerical
simulations as we
have experienced in four-dimensional cases. However, we also conclude
that previous efforts in re-formulating the Einstein equations
can be applied if they are based on constraint propagation analysis.\\
\end{abstract}

\maketitle

\section{Introduction}

Higher dimensional space-time models have been investigated from many
viewpoints in physics.  Current research interests come
from brane-world models that try to solve the hierarchical problem in
the unified theory  (e.g. \cite{Arkani,RS}).
Since these models can be probed by future Large Hadron Collider
experiments, a lot of research is being undertaken.
Even apart from such brane-world models,
many new physical results in higher dimensional general relativity
are reported.  Although we do not have space to list them all,
we mention the discoveries of the black-hole solutions in
five-dimensional space-time (e.g. \cite{BHole5d}) that
violate the traditional black-hole no-hair conjecture,
the possibility of new stable configurations of
black-string models (e.g. \cite{BString5d,BStringNum}), and
the modified version of cosmic hoop conjecture (e.g. \cite{Hoop5d}).

In order to investigate such topics, especially their dynamical
and nonlinear behavior, numerical simulations are necessary.
Numerical relativity is promising research field, but it is
also true that we have not yet obtain the recipe to perform long-term
stable and accurate dynamical evolution.
Many trial simulations of binary compact objects have revealed that
the mathematically equivalent sets of evolution equations
are showing different numerical stability in the free evolution
schemes.  Current research target in numerical relativity is to find
out better reformulation of the Einstein equation (see reviews, e.g.
\cite{Lehner_review,novabook,BaumgarteShapiro_review}).

In this {\it Report},
we study the dimensional dependence of constraint propagation in
the standard Arnowitt-Deser-Misner (ADM) formulation of the
Einstein equation (space-time decomposition) \cite{ADM,ADM-York}.
Reader might think that starting with the ADM equation is old-fashioned
since recent large-scale numerical simulations are not using the ADM
equation due to its stability problem.
However, we still think that ADM is the starting formulation for
analyzing the dynamical behavior both analytically and numerically.
The plenty of re-formulations of the Einstein equations have been
proposed in a last decade. 
Most of them are starting from the ADM variables. 
The practical advantages of such re-formulations are extensively under
investigation by many groups now, but, in our viewpoint,
the essential improvements of them can be explained in a unified way
via constraint propagation equations \cite{novabook}.
As we have shown in \cite{adjADM,adjADM2},
the stability problem of ADM can be controlled by adjusting constraints
appropriately to evolution equations, and that key idea also
works in other formulations \cite{adjAsh,adjBSSN}.
Therefore the analysis of the ADM equation is still essential.

The idea of constraint propagation (originally reported in
\cite{detweiler,Fri-con}) is a useful tool for
calibrating the Einstein equations for numerical simulations.
The modifications to the evolution equations change the property of
the associated constraint propagation, and several particular
adjustments to evolution equations using constraints are expected to
diminish the constraint violating modes.
We proposed to apply eigenvalue analysis to constraint
propagation equations and to judge the property of the constraint
violation.
The proposed adjusted equations have been confirmed as showing better
stability than before by numerical
experiments (e.g. \cite{pennstate,illinois}).
The purpose of this report is to show this idea is also applicable
to all higher dimensional cases.

\section{$N+1$-dimensional ADM equations}
We start from the $(N+1)$-dimensional Einstein equation,
\begin{equation}
G_{\mu\nu}
\equiv {\cal R}_{\mu\nu}-{1\over 2} g_{\mu\nu}{\cal R}
   + g_{\mu\nu} \Lambda =8 \pi T_{\mu\nu},
\label{theEinstein}
\end{equation}
and decompose it into $N$-dimensional space plus time, using
the projection operator $\bot^\mu_\nu$,
\begin{equation}
\gamma^\mu_\nu = \delta ^\mu_\nu + n^\mu n_\nu \equiv \bot^\mu_\nu.
\end{equation}
where $n^\mu$ is a unit normal vector of the spacelike hypersurface
$\Sigma$,
and we write the metric components,
\begin{equation}
ds^2=-\alpha^2 dt^2 + \gamma_{ij}(dx^i+\beta^idt)(dx^j+\beta^jdt),
\end{equation}
where $\gamma_{ij}$ expresses $N$-dimensional intrinsic metric,
and $\alpha$ and $\beta^i$
the lapse and shift function, respectively.
(Greek indices proceed $0, 1, \cdots, N$, while Latin indices
proceed $1, \cdots, N$.)

The projections of the Einstein equation are the following three:
\begin{eqnarray}
G_{\mu\nu} \, n^\mu \,  n^\nu
&=& 8 \pi  \,  T_{\mu\nu} \,  n^\mu \, n^\nu  \equiv 8 \pi \rho_H,
\label{project1}
\\
G_{\mu\nu} \, n^\mu \, \bot_\rho^\nu
&=& 8 \pi   \,  T_{\mu\nu} \, n^\mu
\, \bot_\rho^\nu \equiv - 8 \pi J_\rho,
\label{project2}
\\
G_{\mu\nu} \, \bot_\rho^\mu \, \bot_\sigma^\nu &=& 8 \pi  \,  T_{\mu\nu}
\, \bot_\rho^\mu
\, \bot_\sigma^\nu \equiv   8 \pi S_{\rho\sigma},
\label{project3}
\end{eqnarray}
where we defined
\begin{equation}
T_{\mu\nu}
= \rho_H n_\mu n_\nu + J_\mu n_\nu + J_\nu n_\mu + S_{\mu\nu},
\end{equation}
which gives $T= -\rho_H + S^\ell{}_\ell$.

To express the decomposition,
we introduce the extrinsic curvature $K_{ij}$ as
\begin{eqnarray}
K_{ij}  &\equiv&  -  \bot^\mu_i \bot^\nu_j \nabla_\nu n_{\mu}
\nonumber \\&=&
{ 1\over 2 \alpha}
\left( -  \partial_t \gamma_{ij} + D_j \beta_{i} + D_i \beta_{j}
\right),
\label{gijevo}
\end{eqnarray}
where $\nabla$ and $D_i$ is
the covariant differentiation with respect to
$g_{\mu\nu}$ and $\gamma_{ij}$, respectively.

The projection of the Einstein equation onto $\Sigma$
is given using the Gauss equation,
\begin{equation}
{}^{(N)}R^i_{~jkl}
= {\cal R}^\mu_{~\nu\rho\sigma}
\bot^{~i}_\mu \bot_j^{~\nu} \bot_k^{~\rho}
\bot_l^{~\sigma}
- K^i_{~k}K_{jl}
+ K^i_{~l}K_{jk}\,,
\label{Gauss}
\end{equation}
and the Codazzi equation,
\begin{equation}
D_j K^{~j}_i - D_i K
= - {\cal R}_{\rho\sigma}n^\sigma \bot^{~\rho}_i \,,
\label{Codacci}
\end{equation}
where
$K=K^i{}_i$.
For later convenience, we contract (\ref{Gauss}) to
\begin{eqnarray}
{}^{(N)}R_{ij}
&=& {\cal R}^\mu_{~i \rho j}
(\delta^{~\rho}_\mu + n_\mu n^\rho )
- K K_{i j}
+ K^\ell_{~j}K_{i \ell},
\label{Gauss2}
\\
{}^{(N)}R
&=& {\cal R}
+ 2 {\cal R}_{\mu\nu} n^\mu n^\nu
- K^2
+ K^{ij}K_{ij}.
\label{Gauss3}
\end{eqnarray}

Eqs. (\ref{project1}) and (\ref{Gauss3})
give  the {Hamiltonian constraint}, ${\cal C}_H \approx 0$, where
\begin{eqnarray}
{\cal C}_H & \equiv &
(G_{\mu\nu}-8\pi T_{\mu\nu}) n^\mu n^\nu
\nonumber \\ &=&
{ 1\over 2} ({}^{(N)} R + K^2
- K^{ij}K_{ij})
- 8 \pi \rho_H - \Lambda  ,
\label{constH}
\end{eqnarray}
while (\ref{project2})
and (\ref{Codacci}) give  the {momentum constraint}, ${\cal C}_{Mi}
\approx 0$, where
\begin{eqnarray}
{\cal C}_{Mi} &\equiv&
(G_{\mu\nu}-8\pi T_{\mu\nu}) n^\mu \bot^\nu_i
\nonumber \\&=&
D_j K_i^j - D_i K-  8 \pi  J_i.
\label{constM}
\end{eqnarray}
Both (\ref{constH}) and (\ref{constM}) have the same expression as
those of the four-dimensional version.

The evolution equation for $\gamma_{ij}$ is obtained from
(\ref{gijevo}), which is again the same expression as the four-
dimensional
version.

The evolution equation of $K_{ij}$ is obtained also from
(\ref{project2}).
The contraction of
(\ref{theEinstein}) gives
\begin{eqnarray}
{\cal R}_{ij}&=&
   8\pi \left( S_{ij} -  {1\over N-1}\gamma_{ij} T \right)
-{2 \over 1-N} \gamma_{ij}\Lambda,
\label{contractEinstein}
\end{eqnarray}
where we used $g_{\mu\nu}g^{\mu\nu}=N+1$.
A straightforward calculation of
${\cal R}^\mu{}_{i \rho j}
=
\partial_\rho \Gamma^\mu_{ij} -\partial_j \Gamma^\mu_{\rho i}
+\Gamma^\mu_{\rho \sigma}\Gamma^\sigma_{ij}
-\Gamma^\mu_{\sigma j}\Gamma^\sigma_{i\rho},
$
where $\Gamma^\mu_{\nu\rho}$ is the Christoffel symbol,
gives
\begin{eqnarray}
&&\alpha \, {\cal R}_{\mu i \rho j} n^\mu n^\rho
=
(\partial_tK_{ij})
+(D_jD_i \alpha)
-\beta^k  (D_k K_{ij})
\nonumber \\&&
-(D_j\beta^k)K_{ik}
-(D_i\beta^k)K_{kj}
+\alpha K_{kj}K_i{}^k.
\label{nakamura-a21gy}
\end{eqnarray}
Substituting (\ref{contractEinstein}) and (\ref{nakamura-a21gy}) into
(\ref{Gauss2}),
we obtain
\begin{widetext}
\begin{eqnarray}
\partial_t K_{ij}  &&
=
\alpha{}^{(N)\!}R_{ij}
+ \alpha K K_{ij}
- 2\alpha K^\ell_{~j}K_{i \ell} - D_iD_j \alpha
\nonumber \\&&
+\beta^k  (D_k K_{ij})
+(D_j\beta^k)K_{ik}
+(D_i\beta^k)K_{kj}
\nonumber  \\&&
- 8\pi\alpha \left( S_{ij} -  {1\over N-1}\gamma_{ij} T \right)
-{2 \alpha \over N-1} \gamma_{ij}\Lambda,
\label{KijevoNdim}
\end{eqnarray}
\end{widetext}
that is, only the matter and cosmological constant related terms depend
on the dimension $N$.

If we have matter, we need to evolve the matter thems
together with metric.
The evolution equations for matter terms can be derived from the
conservation
equation,
$\nabla^\mu T_{\mu\nu}=0$.
In the next section, we will discuss them together with
the constraint propagation equations.

\section{$N+1$-dimensional constraint propagation}
The constraint propagation equation,
$\partial_t ({\cal C}_H, {\cal C}_{Mi})^T$,
can be derived in many ways, and among them the derivation via the 
Bianchi
identity \cite{Fri-con} may be the easiest.

In general, we write a $N+1$-dimensional symmetric tensor
${\cal S}_{\mu\nu}$ which obeys the Bianchi identity,
$\nabla^\nu {\cal S}_{\mu\nu}=0$.
Let us express ${\cal S}_{\mu\nu}$ by decomposing as
\begin{equation}
{\cal S}_{\mu\nu}= X n_\mu n_\nu +Y_\mu n_\nu +Y_\nu n_\mu
+Z_{\mu\nu}.
\label{Sdecomp}
\end{equation}
The normal and spatial projections of
$\nabla^\nu {\cal S}_{\mu\nu}$
become
\begin{widetext}
\begin{eqnarray}
n^\mu  \nabla^\nu {\cal S}_{\mu\nu} &=&
-Z_{\mu\nu} (\nabla^\mu n^\nu) 
-\nabla^\mu Y_\mu 
+Y_\nu n^\mu \nabla_\mu n^\nu  
-2Y_\mu n_\nu(\nabla^\nu n^\mu) 
-X(\nabla^\mu n_\mu)  
-n_\mu(\nabla^\mu X), 
\\
h_{i}{}^\mu  \nabla^\nu {\cal S}_{\mu\nu} &=&
\nabla^\mu Z_{i\mu} 
+Y_i(\nabla^\mu n_\mu) 
+Y_\mu(\nabla^\mu n_i) 
+ X (\nabla^\mu n_i)n_\mu 
+n_\mu (\nabla^\mu Y_i), 
\end{eqnarray}
where we used $\nabla$, while \cite{Fri-con} uses different operator.
For convenience, we rewrite them
\begin{eqnarray}
\alpha n^\mu \nabla^\nu {\cal S}_{\mu\nu}
&=&
-(\partial_t X)
+\alpha KX
+\beta^j (\partial_j X)
-\alpha\gamma^{ji}(\partial_i Y_j)
+\alpha(\partial_l \gamma_{mn})(\gamma^{ml}\gamma^{nj}
-(1/2)\gamma^{mn}\gamma^{lj})Y_j
\nonumber \\&&
-2\gamma^{im} (\partial_m\alpha)Y_i
+\alpha K^{ij} Z_{ij},
\label{bianchiprop1}
\\
-\alpha h^{\mu}_i \nabla^\nu {\cal S}_{\mu\nu}
&=&
- (\partial_t Y_i)
-(\partial_i\alpha) X
+\alpha K Y_i
+\beta^j (\partial_j Y_i)
+\gamma^{km}(\partial_i\beta_m) Y_k
-\beta^j\gamma^{kp}(\partial_i\gamma_{pj})Y_k
\nonumber \\&&
-\alpha\gamma^{jk}(\partial_k Z_{ij})                
-(\partial_j\alpha)Z_{i}{}^j     
+(1/2)\alpha(\partial_i\gamma_{jk})Z^{kj}            
+\alpha\gamma^{mk}(\partial_m\gamma_{kj})Z_{i}{}^j      
-(1/2)\alpha\gamma^{mk}(\partial_j\gamma_{mk})Z_{i}{}^j,  
\label{bianchiprop2}
\end{eqnarray}
respectively.
\end{widetext}

If we substitute
$({\cal S}_{\mu\nu}, X, Y_{i}, Z_{ij})=(T_{\mu\nu}, \rho_H, J_i, 
S_{ij})$
into (\ref{bianchiprop1}) and (\ref{bianchiprop2}) and assume
$\nabla^\mu T_{\mu\nu}=0$,  then
we obtain the matter evolution equations, $\partial_t \rho_H$ and
$\partial_t J_i$.
If we substitute
$({\cal S}_{\mu\nu}, X, Y_{i}, Z_{ij})=
(G_{\mu\nu}-8\pi T_{\mu\nu}, {\cal C}_H, {\cal C}_{Mi},
\kappa \gamma_{ij}{\cal C}_H)$ with $\kappa$=const. and assume
$\nabla^\mu (G_{\mu\nu}-8\pi T_{\mu\nu})=0$ , then
we obtain the constraint propagation equations,
$\partial_t {\cal C}_H$ and $\partial_t {\cal C}_{Mi}$.
[The parameter $\kappa$ corresponds to
adding a term to (\ref{KijevoNdim}),
$+(\kappa-1) {\cal C}_H$. ]

This derivation does not depend on the dimension $N$ at all.
Therefore the evolution equations both for the matter and constraints
remain the same with those in the traditional four dimensional version.

The constraints include the extrinsic curvature terms, and the evolution
equation of $K_{ij}$ changes due to $N$ as we saw in
(\ref{KijevoNdim}).
Interestingly, however, such changes will be cancelled out and the
resultant constraint propagation equations remain the same.

This means that a series of
constraint propagation analyses can be directly applied to
higher dimensional space-time.
That is, the standard ADM evolution equations are likely to fail
for long-term stable simulations.
However, previously proposed adjustment techniques
(e.g. \cite{adjADM,adjADM2}) are also effective.

For example, constraint amplification factors (i.e. the eigenvalues
of constraint propagation matrix) in Schwarzschild space-time
[eq. (47)  in \cite{adjADM2}] are $(0,0, \pm f(r))$
for four-dimensional standard ADM evolution equations,
where $f(r)$ is a complex-valued function.
In the five-dimensional
Schwarzschild or black-string
case, they become simply $(0,0,0, \pm f(r))$.

\section{Remarks}
Motivated by the recent interests in higher dimensional space-time,
we checked the constraint propagation equations based on the
$N+1$ ADM scheme.
The evolution equation has matter terms which depend on $N$,
but we show the constraint propagations remain the same as those in the
four-dimensional ones.
This indicates that there would be problems with accuracy and stability
when we directly apply the $N+1$ ADM formulation to numerical 
simulations
as we have experienced in four-dimensional cases. However, we also
conclude that previous efforts in re-formulating the Einstein equations
can be applied if they are based on constraint propagation analysis.
The generality holds for other systems when their 
constraints are writtin in the form of (\ref{Sdecomp}).

Since we only used the Bianchi identity in the core discussion,
the assertion is also applicable to brane-world models.
In the context of the Randall-Sundrum brane-world models \cite{RS},
people study the modified four-dimensional Einstein equations
\cite{ShiromizuMaedaSasaki},
which are derived from five-dimensional Einstein equations with
a thin-shell (3-brane) approximation.
The terms there additional to the standard ADM
(see eq.(17) in \cite{ShiromizuMaedaSasaki}) include
extrinsic curvature (due to shell-normal vector), cosmological
constant(s), and five-dimensional Weyl curvature.
These terms, however, can be interpreted as a single stress-energy
tensor which obeys the Bianchi identity.  Therefore the properties
of the constraint propagation equations are the same as the above
(from the five-dimensional space-time viewpoint).
Our proposals for the adjustments \cite{adjADM,adjADM2} are
also valid in brane-world models.

We hope this short report helps numerical relativists for developing
their future simulations.

{\bf Acknowledgements} ~~
HS is supported by the special postdoctoral researchers' program
at  RIKEN.
This work was supported partially by the Grant-in-Aid for Scientific
Research Fund of the Japan Society of the Promotion of Science, No.
14740179
and by Waseda University Grant for Special Research Projects 
Individual Research 2003A-870.



%
\end{document}